\documentclass[english]{IEEEtran}
\usepackage[T1]{fontenc}
\usepackage[latin9]{inputenc}
\usepackage{amsmath}
\usepackage{graphicx}
\usepackage{esint}
\usepackage{babel}
\begin{document}

\title{Stability and Hopf Bifurcation Analysis of the \\
Delay Logistic Equation}

\author{Milind M Rao$^{\text{\#}}$ and Preetish K L$^{\text{*}}$\\
$^{\text{\#}}$Department of Electrical Engineering, IIT Madras, Chennai-600036,
India\\
$^{\text{*}}$Department of Mechanical Engineering, IIT Madras, Chennai-600036,
India\\
E-mail: milindmrao@gmail.com and kl.preetish@gmail.com}
\maketitle
\begin{abstract}
Logistic functions are good models of biological population growth.
They are also popular in marketing in modelling demand-supply curves
and in a different context, to chart the sales of new products over
time. 

Delays being inherent in any biological system, we seek to analyse
the effect of delays on the growth of populations governed by the
logistic equation. In this paper, the local stability analysis, rate
of convergence and local bifurcation analysis of the logistic equation
with one and two delays is carried out and it can be extended to a
system with multiple delays.

Since fluctuating populations are susceptible to extinction due to
sudden and unforeseen environmental disturbances, a knowledge of the
conditions in which the population density is fluctuating or stable
is of great interest in planning and designing control as well as
management strategies. \end{abstract}
\begin{IEEEkeywords}
delay logistic equation, stability, rate of convergence, Hopf bifurcation
analysis.
\end{IEEEkeywords}

\section{Introduction}

The logistic equation is a simple model of population growth in conditions
where there are limited resources. It was proposed by Verhulst in
1838 to describe the self limiting growth of a biological population
{[}1{]}. The equation has a variety of applications. It is used in
neural networks to clamp signals to within a specified range {[}2{]},
in economics to illustrate the progress of the diffusion of an innovation
through its life cycle {[}3{]}, in medicine to model the growth of
tumours {[}4{]} and in linguistics to model language change {[}5{]}.

The growth of a population is conceptualised in Fig. 1. A biological
population with plenty of food, space to grow and no threat from predators
tends to grow at a rate that is proportional to the population. However,
most populations are constrained by environmental limitations. Growth
is eventually limited by a factor, usually one from amongst many essential
resources. When a population is far from its limits of growth or the
carrying capacity of the ecosystem, it can grow exponentially. The
feedback about the availability of resources reaches with a delay
due to various factors such as generation and maturation periods,
differential resource consumption with respect to age structure, hunger
threshold levels, migration and diffusion of populations, markedly
differing birth rates in interaction species and delays in behavioural
responses to a changing environment (including changes in density
of prey or predators or competing species) {[}6-8{]}. When nearing
its limits, the population can fluctuate, even chaotically. 

According to the logistic equation, the growth rate of a population
is directly proportional to the current population and the availability
of resources in the ecosystem. The logistic equation model is as follows,

\begin{equation}
\frac{dy}{dt}=ay\left(1-\frac{y}{k}\right),
\end{equation}
where $y$ is the population at that instant, $a$ is called the Malthusian
parameter which represents the growth rate and $k$ is the carrying
capacity of the ecosystem.

Hutchinson incorporated the effect of delays into the logistic equation
{[}9{]}. Delays bring about interesting topological changes in the
population size like damped oscillations, limit cycles and even chaos
{[}10{]}. The bifurcation analysis of a system with a single delay
has been performed in {[}11{]}. The importance of two delays in the
logistic equation can be seen in {[}12-15{]}. In this paper, the effect
of such delays will be analysed. Analysis will primarily focus on
local stability, local rate of convergence and local bifurcation phenomena.
The procedure adopted here to characterise a system with two delays
can be easily extended to a case of multiple delays.

The rest of the paper is organized as follows. In section 2, the model
is described. In section 3, It is linearised, local stability analysis
of the model is carried out and the conditions for stability are presented.
In section 4, rate of convergence is analysed. Bifurcation analysis
of the system is carried out in section 5. Finally in section 6, the
graphical results of the analysis are presented.

\begin{figure}[h]
\includegraphics[width=8cm]{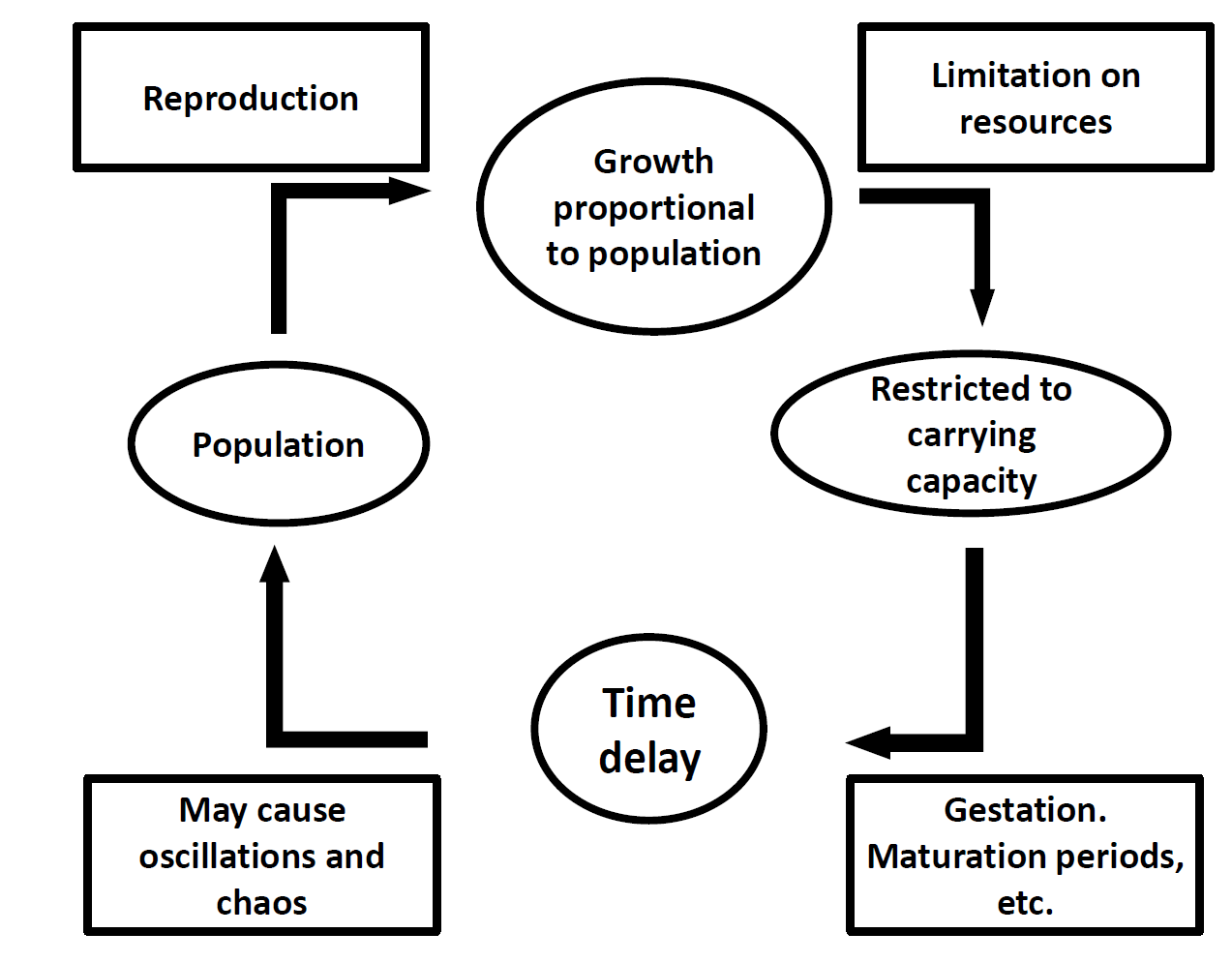}

\caption{Schematic of the model. The population growth depends both on the
current population (reproductive growth) as well as the availability
of resources. The logistic equation with delay model the abstraction
well and is widely used. }
\end{figure}

\section{Model Description}

In this section, we present the model employed with the accompanying
assumptions for a single delay system and a two delay system.

\subsection{Logistic equation with single delay}

The model we have employed to demonstrate population dynamics is as
follows:
\begin{itemize}
\item The initial normalised population is chosen to be small (typically
0.01) as it cannot be zero. A zero initial population signifies a
non-existent species. 
\item The equation is normalized. 
\item The growth rate is $a$ which is finite, positive and time independent. 
\item The delay $\tau_{1}$ is also finite and positive. 
\end{itemize}
With these assumptions, the single delay logistic equation simplifies
to

\begin{equation}
\frac{dx}{dt}=ax(t)\left[1-bx(t-\tau_{1})\right].
\end{equation}

\subsection{Logistic equation with multiple delays}

The same assumptions as made is the single delay logistic equation
my be applied here as well. In the case of a system with $n$ delays,
the equation takes the form

\begin{equation}
\frac{dx}{dt}=ax(t)\left[1-\overset{n}{\underset{i=1}{\sum}}b_{i}x(t-\tau_{i})\right].
\end{equation}
The case of two delays is analysed in this paper.

\section{Local Stability Analysis}

The delay logistic equation is non-linear. In this section, we first
linearise it and proceed to extract conditions for the stability of
both the single delay as well as the multiple delay logistic equation.
We end the section by presenting stability charts of the system.

\subsection{Stability analysis of logistic equation with single delay}

Let the equilibrium point (where the growth rate is zero) be denoted
by $x^{*}$. Then $x(t)=x(t-\tau_{1})=x^{*}$. Also,

\begin{equation}
\frac{dx}{dt}\equiv0=ax^{*}(1-bx^{*}).
\end{equation}
Solving for this, we get $x^{*}=0$ (signifying the initial stages)
and $x^{*}=\frac{1}{b}$ (signifying the saturating stages).

Now, to linearise the equation, substitute $x(t)\equiv x^{*}+p(t)$,
where $p(t)$ is a small variation in the population. Higher powers
of $p$ may be neglected. The equation therefore is

\[
\frac{dx}{dt}=a\left[x^{*}+p(t)\right]\left[1-bx^{*}-bp(t-\tau_{1})\right].
\]

\subsubsection{Stability analysis around initial value}

Substituting $x^{*}=0$ or linearising around the initial value, we
have 

\begin{equation}
\frac{dp}{dt}=ap(t),
\end{equation}
the solution of which is easily obtained as

\[
p(t)=b_{1}e^{at},
\]
which is to say

\begin{equation}
x(t)=be^{at},
\end{equation}
around $x=0$. This equilibrium point is not stable as the characteristic
equation of ($5)$ has as its root $a$ which is in the right half
plane.

\subsubsection{Stability analysis around saturation value}

This value is of greater interest to us. Substituting $x^{*}=\frac{1}{b}$
or linearising around the saturation value, we have 

\[
\frac{dp}{dt}=-ap(t-\tau_{1}).
\]
To solve this, we take

\[
p(t)=b_{2}e^{\lambda t}.
\]
Then, the characteristic equation reduces to 

\begin{equation}
\lambda+ae^{-\lambda\tau_{1}}=0.
\end{equation}
For the system to be stable, the roots of the above equation must
lie on the left-half of the $\lambda$ plane. Bifurcation point is
the value of the parameters for which the roots lie on the imaginary
axis.

Substituting $\lambda=j\omega$ in $(7)$, we get

\[
j\omega+ae^{-j\omega\tau_{1}}=0
\]

\begin{equation}
\Longrightarrow j\omega+a\left[\cos(\omega\tau_{1})-j\sin(\omega\tau_{1})\right]=0.
\end{equation}
Equating real part to zero, we get

\[
a\cos(\omega\tau_{1})=0
\]

\begin{equation}
\Longrightarrow\omega\tau_{1}=(2n+1)\frac{\pi}{2},n=0,1,2...
\end{equation}
Equating imaginary part of $(8)$ to zero, we get

\[
\omega-a\sin(\omega\tau_{1})=0.
\]
To get minimum order solution, we use $n=0$ in $(9)$.

\[
\frac{\pi}{2}-a\tau_{1}=0.
\]
For $\tau=0$, the characteristic equation is 

\[
\lambda=-a,
\]
which implies that $\lambda<0$ for $a>0$ and the system is stable.
i.e., for $a\tau_{1}=0$, the roots of the characteristic equation
lie of the left-half plane. Therefore, necessary and sufficient condition
for stability is,

\begin{equation}
a\tau_{1}<\frac{\pi}{2}.
\end{equation}
It is apparent that if the system has no delays or $\tau_{1}=0$,
the system is always stable.

\subsection{Stability analysis of logistic equation with two delays}

At the fixed points, $x(t-\tau_{1})=x(t-\tau_{2})=x(t)=x^{*}$. Here
$x^{*}$is the equilibrium point. The delay logistic equation then
becomes:

\begin{equation}
\frac{dx}{dt}\equiv0=ax^{*}\left[1-(b_{1}+b_{2})x^{*}\right].
\end{equation}
Equating $\frac{dx}{dt}$ to zero, we get $x*$ to be $0$ or $\frac{1}{b_{1}+b_{2}}$.
We have seen in the previous subsection that the first equilibrium
is unstable and is not of much interest. We explore the latter value.

The delay logistic equation is then linearised by expanding it by
a Taylor series and neglecting all higher order terms. We obtain the
following linearised equation about the fixed point $x*=\frac{1}{b_{1}+b_{2}}$
, such that $x(t)=x^{*}+y(t)$. We obtain:

\begin{equation}
\frac{dy}{dt}=-ax^{*}\left[b_{1}y(t-\tau_{1})+b_{2}y(t-\tau_{2})\right].
\end{equation}
To solve this, we take

\[
y(t)=b_{3}e^{\lambda t}.
\]
If $b_{1}\gg b_{2}$, then the second delay term in ($12$) can be
neglected and the analysis reduces to the case of a single delay system.
Consider the case when $b_{1}=b_{2}$,

\[
\lambda e^{\lambda t}=-\frac{a}{2}\left(e^{\lambda(t-\tau_{1})}+e^{\lambda(t-\tau_{2})}\right).
\]
The characteristic equation therefore is

\begin{equation}
\lambda+\frac{a}{2}e^{-\lambda\tau_{1}}+\frac{a}{2}e^{-\lambda\tau_{2}}=0.
\end{equation}
For the system to be stable, the roots of the above equation must
lie on the left-half of the $\lambda$ plane. Bifurcation point is
the value of the parameters for which the roots lie on the imaginary
axis. Substituting $\lambda=j\omega$ in $(13)$, we get

\[
j\omega+\frac{a}{2}e^{-j\omega\tau_{1}}+\frac{a}{2}e^{-j\omega\tau_{2}}=0
\]

\begin{eqnarray*}
\Longrightarrow j\omega+\frac{a}{2}\left[\cos(\omega\tau_{1})-j\sin(\omega\tau_{1})\right]\\
+\frac{a}{2}\left[\cos(\omega\tau_{2})-j\sin(\omega\tau_{2})\right] & = & 0.
\end{eqnarray*}
Equating imaginary part of $(13)$ to zero, we get

\[
\omega-\frac{a}{2}\sin(\omega\tau_{1})-\frac{a}{2}\sin(\omega\tau_{2})=0.
\]
Equating real part to zero, we get

\[
\frac{a}{2}\cos(\omega\tau_{1})+\frac{a}{2}\cos(\omega\tau_{2})=0
\]

\[
\Longrightarrow\cos\left(\omega\frac{\tau_{1}+\tau_{2}}{2}\right)\cos\left(\omega\frac{\tau_{1}-\tau_{2}}{2}\right)=0.
\]
The second term cannot be $0$ as the condition on the imaginary component
will imply that $\omega=0$ which is not true for all $\omega$.

\begin{equation}
\omega=\frac{2(\pm\frac{\pi}{2}+2\pi C_{1})}{\tau_{1}+\tau_{2}},
\end{equation}
where $C_{1}\in Z$. 

Using the value of $\omega$ obtained in ($14$), we get

\begin{equation}
a=\frac{2\pi}{(\tau_{1}+\tau_{2})\left(\sin\left(\frac{\pi\tau_{1}}{\tau_{1}+\tau_{2}}\right)-\sin\left(\frac{\pi\tau_{2}}{\tau_{1}+\tau_{2}}\right)\right)},
\end{equation}
where $(\tau_{1}+\tau_{2})\left[\sin(\frac{\pi\tau_{1}}{\tau_{1}+\tau_{2}})-\sin(\frac{\pi\tau_{2}}{\tau_{1}+\tau_{2}})\right]\neq0$. 

On simplifying this, we get the bifurcation point at 

\begin{equation}
a=\frac{\pi}{(\tau_{1}+\tau_{2})\cos\left(\frac{\pi(\tau_{1}-\tau_{2})}{2(\tau_{1}+\tau_{2})}\right)}.
\end{equation}
The sufficient and necessary condition is:

\begin{equation}
a(\tau_{1}+\tau_{2})\cos\left(\frac{\pi(\tau_{1}-\tau_{2})}{2(\tau_{1}+\tau_{2})}\right)<\pi.
\end{equation}
Substituting $\tau_{1}=\tau_{2}=\tau$, we get the familiar single
delay case $a\tau<\frac{\pi}{2}$. These conditions are valid for
any positive value of $a,\tau_{1}$and $\tau_{2}$. We also get the
following sufficient condition for the specific case of $b_{1}=b_{2}$
from ($17$),

\[
a(\tau_{1}+\tau_{2})<\pi.
\]

\subsection{Sufficient conditions for stability}

In this sub-section, we explore the sufficient conditions for stability
in both the single as well as multiple delay models. Consider the
characteristic equation of a system with two delays:

\[
\lambda+ax^{*}b_{1}e^{-\lambda\tau_{1}}+ax^{*}b_{2}e^{-\lambda\tau_{2}}=0.
\]
This can be re-written as follows {[}16{]}, {[}17{]}:

\begin{eqnarray*}
\lambda+ax^{*}b_{1}+ax^{*}b_{2}+\frac{\lambda\tau_{1}ax^{*}b_{1}}{\lambda\tau_{1}}\left(e^{-\lambda\tau_{1}}-1\right)\\
+\frac{\lambda\tau_{2}ax^{*}b_{2}}{\lambda\tau_{2}}\left(e^{-\lambda\tau_{2}}-1\right) & = & 0.
\end{eqnarray*}
We now define 

\[
H_{1}(\lambda)=\lambda+ax^{*}b_{1}+ax^{*}b_{2},
\]
and

\[
H_{2}(\lambda)=\lambda\tau_{1}ax^{*}b_{1}\left(\frac{e^{-\lambda\tau_{1}}-1}{\lambda\tau_{1}}\right)+\lambda\tau_{2}ax^{*}b_{2}\left(\frac{e^{-\lambda\tau_{2}}-1}{\lambda\tau_{2}}\right).
\]
Clearly, $H_{1}(\lambda)$ has no zeros on the right half plane. Consider
the imaginary axis,

\[
\mid H_{1}(\lambda)\mid>\mid\lambda\mid.
\]
For any real $\theta,$ 
\[
|\frac{1-e^{-j\theta}}{j\theta}|<1.
\]
Hence, on the imaginary axis we have, 
\[
\mid H_{2}(\lambda)\mid<\mid\lambda\mid\mid\tau_{1}ax^{*}b_{1}+\tau_{2}ax^{*}b_{2}\mid.
\]
If 
\begin{equation}
\tau_{1}ax^{*}b_{1}+\tau_{2}ax^{*}b_{2}<1,
\end{equation}
then:

\[
\mid H_{2}(\lambda)\mid<\mid\lambda\mid.
\]
Hence, by Rouche's Theorem, $H_{1}(\lambda)+H_{2}(\lambda)\neq0$
on the imaginary axis. There cannot be any zeros in the right half
plane and the system is stable. Hence a sufficient condition for stability
is:

\begin{equation}
\frac{a\left(b_{1}\tau_{1}+b_{2}\tau_{2}\right)}{b_{1}+b_{2}}<1.
\end{equation}
For a system with only one delay. i.e. $b_{2}=0$, the sufficient
condition is

\begin{equation}
a\tau_{1}<1.
\end{equation}
It is noted that the condition ($19$) is conservative.

\subsection{Stability charts}

In Fig. $2$, the stability chart of a system with single delay is
seen. The plot simply corresponds to the bifurcation point condition
$a\tau=\frac{\pi}{2}$. Fig. $3$ has the stability chart of the two
delay system. Fig. $4$, the stability of the two delay logistic equation
is shown with respect to the growth rate and the two delays. 

\begin{figure}
\includegraphics[width=8cm]{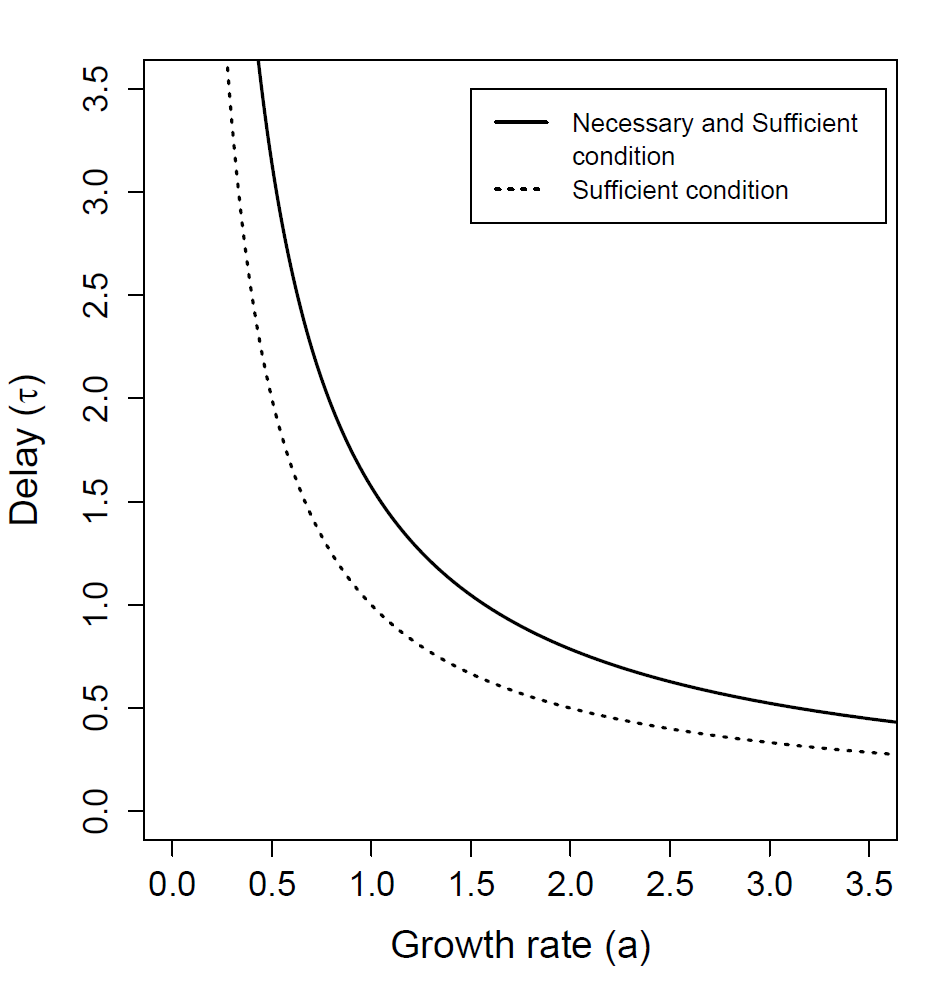}

\caption{Stability chart of a single delay system. Neither the feedback delay
nor the growth rate must be too large for the system to be stable.
The region that is stable is below the necessary and sufficient condition
and Hopf bifurcation occurs on this line.}
\end{figure}

\begin{figure}
\includegraphics[width=8cm]{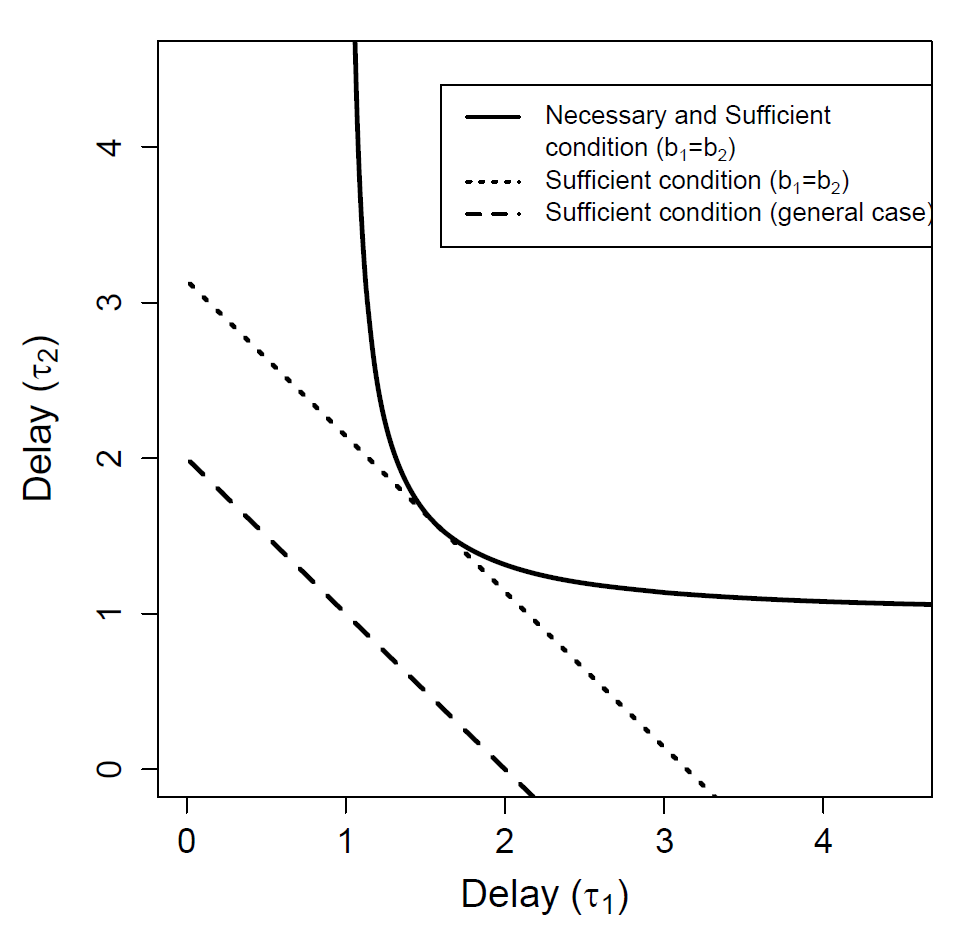}

\caption{Stability chart of a two delay system. Parameter $a=1$. The region
below the curve is stable and corresponds to the necessary and sufficient
condition for stability. }
\end{figure}

\begin{figure}
\includegraphics[width=8cm]{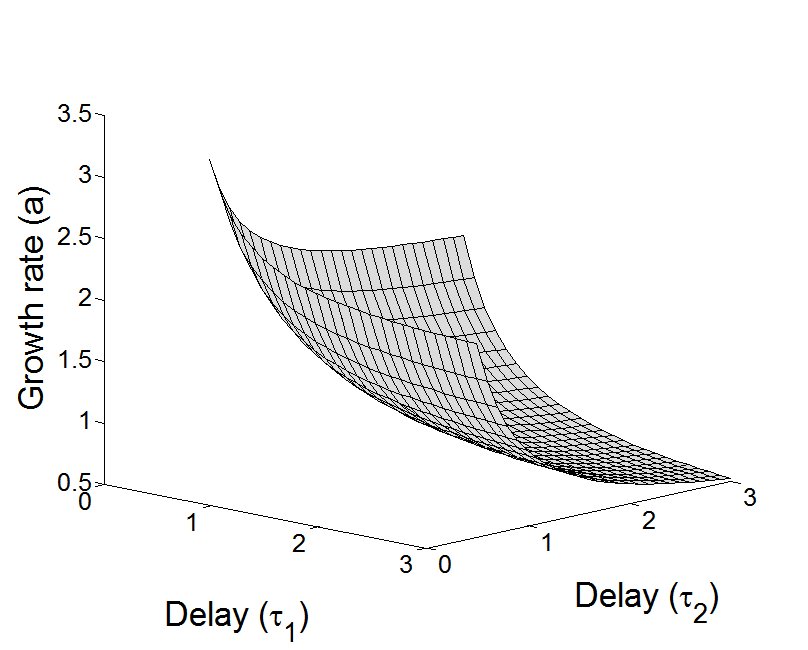}

\caption{Stability chart of a two delay system. The system is stable below
the surface and enters a Hopf bifurcation at the contour shown. It
is not locally stable for points which lie above the contour. }
\end{figure}

\subsection{Nyquist Plots}

In this subsection, we see a few representative Nyquist plots of the
system when stable or unstable.

\subsubsection{Single delay system}

In Fig. $5$, we observe the Nyquist plot of the characteristic equation
($7$). If there are any encirclements about the origin, the system
is unstable as the characteristic equation does not have poles.

\begin{figure}
\includegraphics[width=8cm]{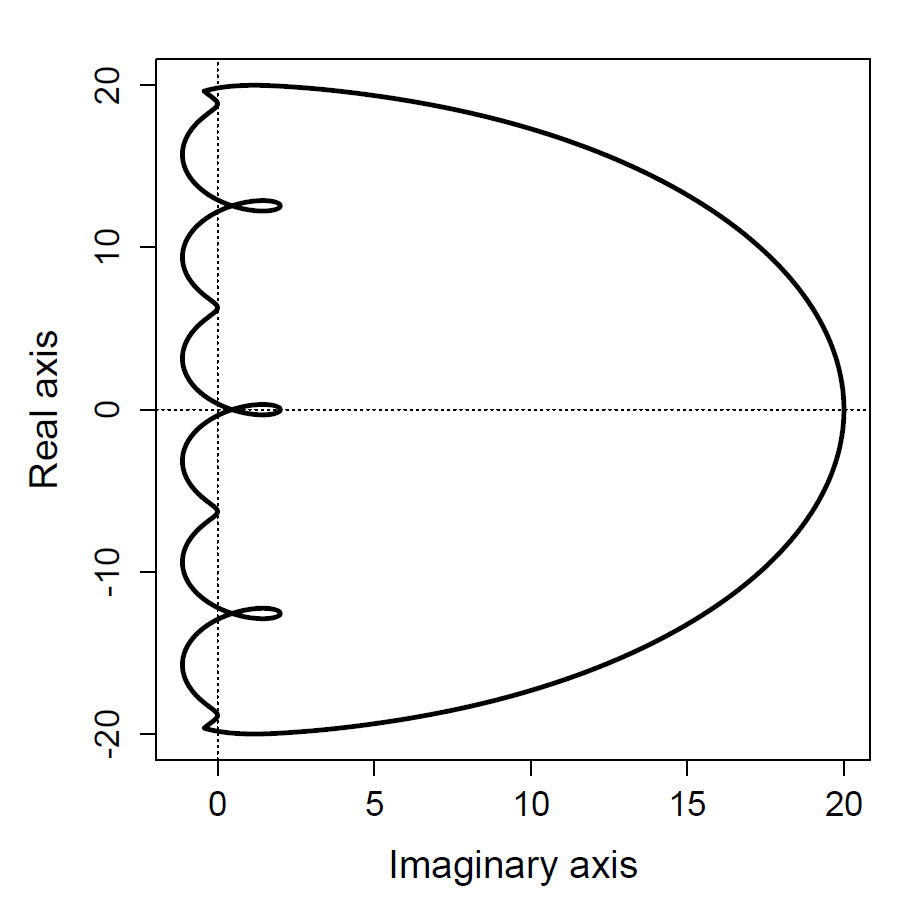}

\caption{Nyquist plot for an unstable system with a single delay. $a=2,\tau=1$
in this diagram and it does not satisfy the sufficient and necessary
conditions highlighted in ($10$). Encirclements are seen about the
origin.}
\end{figure}

\subsubsection{Two delay system}

In Fig. $6$, the Nyquist plot for a system with two delays is shown.
There are no encirclements about the origin, hence the system is stable.
This matches with our calculation as the parameters satisfy the conditions
in ($17$).

\begin{figure}
\includegraphics[width=8cm]{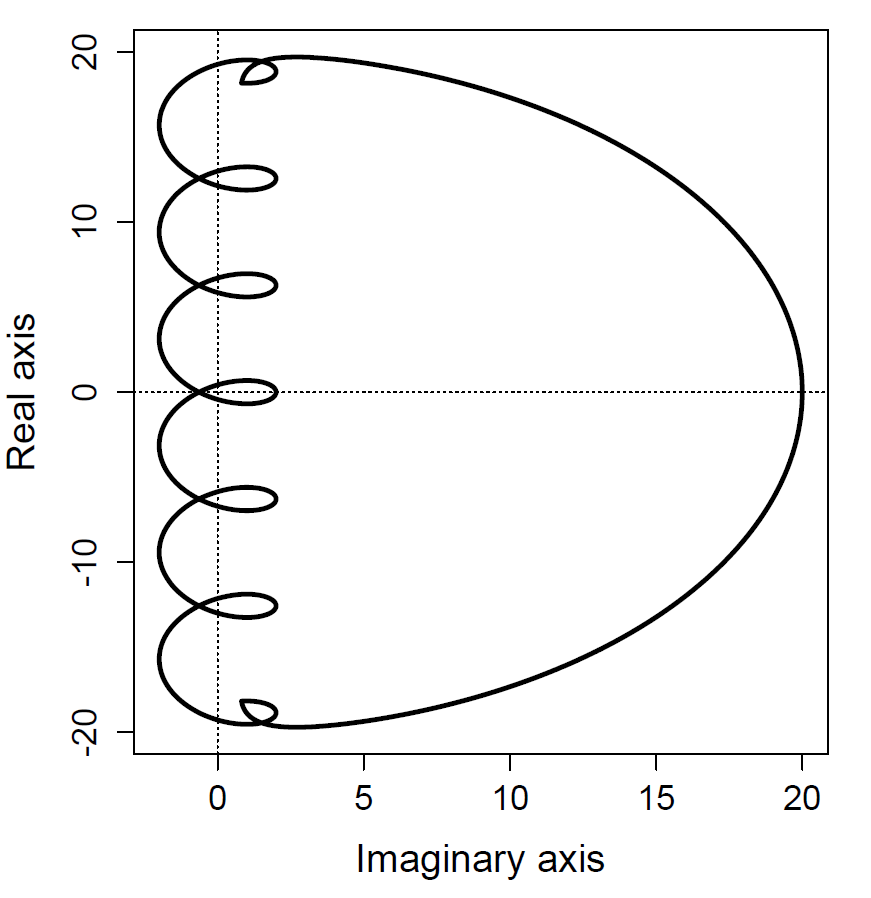}

\caption{Nyquist plot of a system with two delays. $a=b_{1}=b_{2}=1$. $\tau_{1}=0.5$
and $\tau_{2}=1$ is chosen such that conditions in ($17$) are met.
There are no encirclements about the origin and the system is stable.}
\end{figure}

\section{Rate of Convergence}

In this section, the analysis of rate of convergence is carried out
for a system with a single delay. We define the rate of convergence
($R)$ as the inverse of the settling time based on a tolerance band
of $\pm36.8\%$.

Consider the characteristic equation ($7$). If there are no oscillations
in the system, the imaginary part of $\lambda$ is zero. Let $\Re\left\{ \lambda\right\} =\sigma$,
where $\Re(x)$ denotes the real part of $x$. The equation now becomes, 

\begin{equation}
\sigma+ae^{-\sigma\tau}=0.
\end{equation}
On differentiating ($21$) with respect to $\sigma$, we obtain,

\begin{equation}
1-a\tau e^{-\sigma\tau}=0.
\end{equation}
Eliminating $\sigma$ from ($21$) and ($22$), we find,

\begin{equation}
a\tau=\frac{1}{e}.
\end{equation}
This point can be considered as the critical damping point. For values
of $a\tau<\frac{1}{e},$the system behaves in an overdamped fashion
and converges without any overshoot. When $\frac{1}{e}<a\tau<\frac{\pi}{2}$,
underdamped behaviour is observed. i.e. convergent oscillations are
present. 

The characteristic equation ($7$) can be rewritten as in {[}$18${]},

\[
\lambda\tau_{1}e^{\lambda\tau_{1}}=-a\tau_{1}
\]

\[
\Longrightarrow\lambda=\frac{W(-a\tau_{1})}{\tau_{1}},
\]
where $W(x)$ is the Lambert W function. The rate of convergence $R$
is given by,

\begin{equation}
R=\left|\Re\left(\frac{W(-a\tau_{1})}{\tau_{1}}\right)\right|.
\end{equation}

\subsection*{Rate of convergence charts}

In Fig. $7$, the rate of convergence of the logistic equation with
single delay is seen with respect to the time delay. It can be seen
that the system converges fastest for a slightly underdamped system.
Fig. $8$ shows the variation of rate of convergence of solutions
to ($7$) with respect to the growth rate. It can be observed that
rate of convergence tends to zero as growth rate tends to zero. In
other words, a population which multiplies very slowly requires a
very large time to reach its carrying capacity. Fig. $9$ shows the
variation of the rate of convergence with both parameters of the single
delay system.

\begin{figure}
\includegraphics[width=8cm]{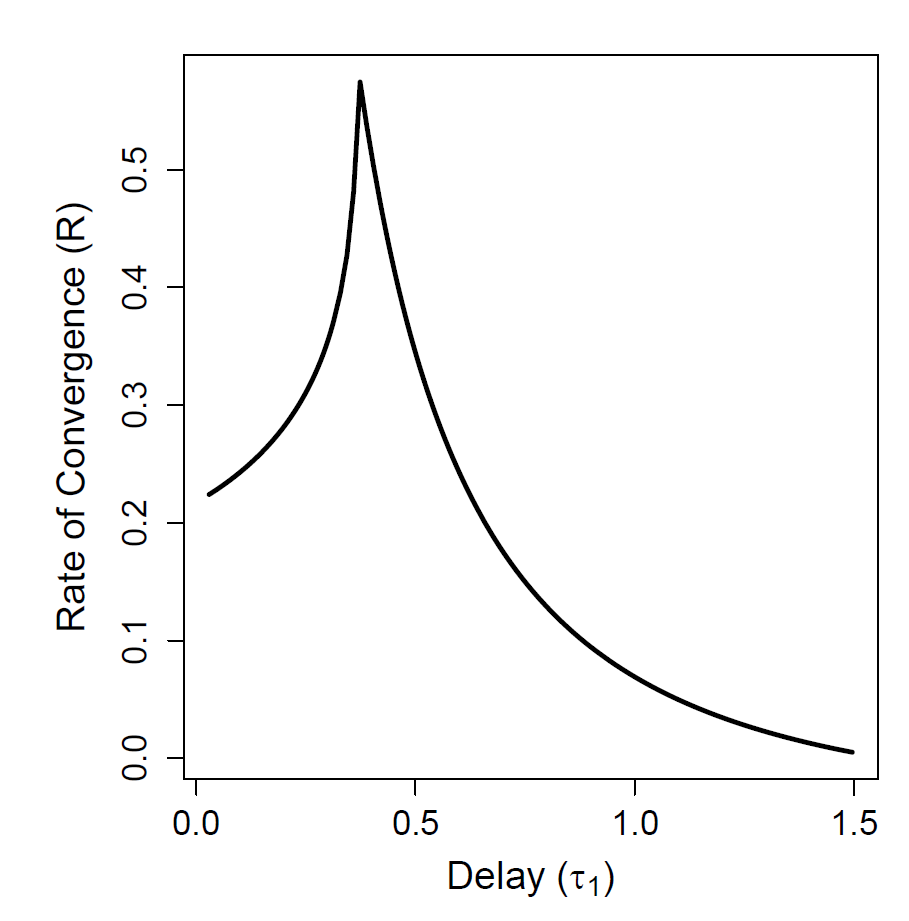}

\caption{The rate of convergence is plotted with respect to the time delay
in a system with single delay where the growth rate $a=1$. It is
seen to converge fastest for a slightly underdamped system when $\tau_{1}$is
slightly greater than the critical damping point $\tau_{1}=\frac{1}{e}$.}
\end{figure}

\begin{figure}
\includegraphics[width=8cm]{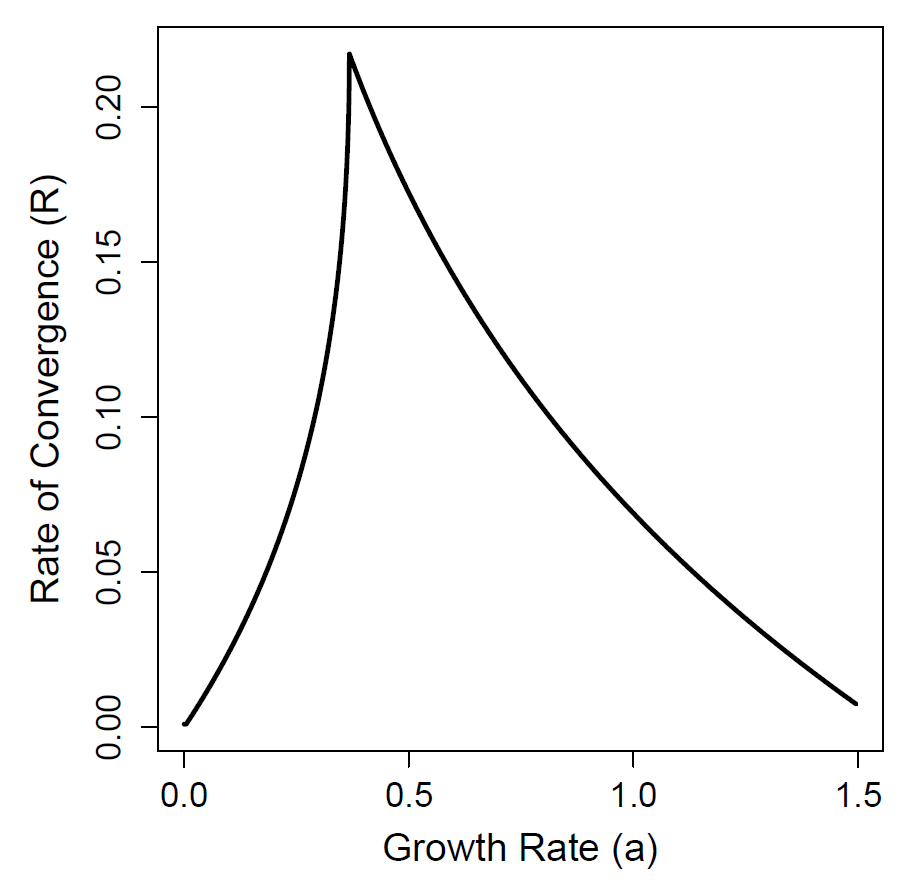}

\caption{The rate of convergence is plotted with respect to the time delay
in a system with single delay where the delay $\tau_{1}$ = 1.}
\end{figure}

\begin{figure}
\includegraphics[width=8cm]{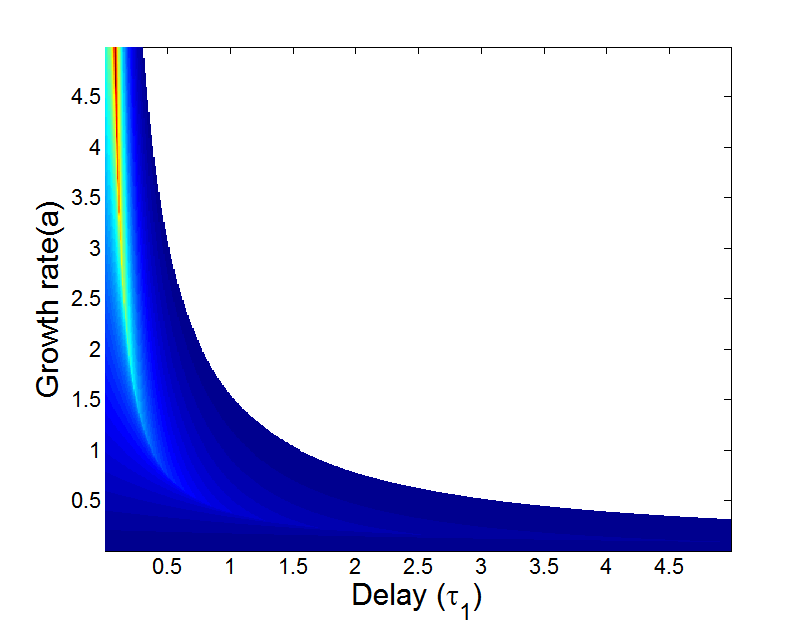}

\caption{Rate of convergence with respect to growth rate $a$ and delay $\tau_{1}$
in the single delay system. The red portions represent the fastest
rate of convergence and blue the slowest. The white portions are for
the regions where the system is not stable. As can be seen from the
chart, fastest rates of convergence are obtained for systems with
large values of growth rate and a small value of the delay.}
\end{figure}

\section{Bifurcation Analysis}

In this section, it is shown that the delay logistic equation undergoes
a Hopf bifurcation at a critical value of the parameters growth rate
($a$) or delay ($\tau_{1}$). The nature of the bifurcation is characterized
in the subsequent sub-sections as are the properties of the resulting
periodic oscillations. Bifurcation diagrams are presented at the end
of the section.

\subsection{Existence of Hopf bifurcation}

From ($10$), at the critical point for the single delay case, $a\tau_{1}=\frac{\pi}{2}$.
Differentiating characteristic equation ($7$) with respect to the
growth rate, we get: 

\[
\frac{d\lambda}{da}=\frac{e^{-\lambda\tau_{1}}}{a\tau_{1}e^{-\lambda\tau_{1}}-1}.
\]
Evaluating at critical point $a^{*}=\frac{\pi}{2\tau_{1}},$

\begin{equation}
\Re\left\{ \frac{d\lambda}{da}\right\} _{a=a^{*}}=\frac{\frac{\pi}{2}}{\frac{\pi^{2}}{4}+1}>0.
\end{equation}

Similarly differentiating ($7$) with respect to the delay , we obtain: 

\[
\frac{d\lambda}{d\tau_{1}}=\frac{\lambda ae^{-\lambda\tau_{1}}}{1-a\tau_{1}e^{-\lambda\tau_{1}}}.
\]
Evaluating at critical point $\tau_{1}^{*}=\frac{\pi}{2a},$

\begin{equation}
\Re\left\{ \frac{d\lambda}{d\tau_{1}}\right\} _{\tau_{1}=\tau_{1}^{*}}=\frac{a^{2}}{\frac{\pi^{2}}{4}+1}>0.
\end{equation}

The transversality condition of the Hopf spectrum with respect to
the growth rate and time delay is satisfied in ($25$) and ($26$)
respectively. Thus, logistic equation with single delay undergoes
a Hopf bifurcation at the critical point given by $a\tau_{1}=\frac{\pi}{2}.$ 

Consider the logistic equation with two delays given by ($13$). Differentiating
($13$) with respect to the growth rate we get,

\[
\frac{d\lambda}{da}=\frac{e^{-\lambda\tau_{1}}+e^{-\lambda\tau_{2}}}{a\tau_{1}e^{-\lambda\tau_{1}}+a\tau_{2}e^{-\lambda\tau_{2}}-2}.
\]
Evaluating at the critical point given by 

\begin{equation}
a^{*}=\frac{\pi}{(\tau_{1}+\tau_{2})\cos\left(\frac{\pi(\tau_{1}-\tau_{2})}{2(\tau_{1}+\tau_{2})}\right)},
\end{equation}
we get,

\begin{equation}
\Re\left\{ \frac{d\lambda}{da}\right\} _{a=a^{*}}=\frac{a(\tau_{1}+\tau_{2})+\pi}{a^{2}\tau_{1}^{2}+a^{2}\tau_{2}^{2}+\frac{2\pi a\tau_{1}\tau_{2}}{\tau_{1}+\tau_{2}}}>0.
\end{equation}

Hence the transversality condition of the Hopf bifurcation is satisfied
with respect to the growth rate. Thus, the two delay logistic equation
undergoes Hopf bifurctaion with respect to the growth rate at the
critical point given by ($27$).

\subsection{Direction and stability of the Hopf bifurcation in the logistic equation
with growth rate as the parameter}

The logistic equation with a single delay or with two delays undergoes
a Hopf at a critical value of the growth rate as shown in the previous
sub-section. In this section the direction, stability and period of
the bifurcating solutions is analysed. The procedure adopted is based
on the centre manifold theory {[}19{]} (See also {[}20{]}). The logistic
equation with two delays as described in (3) can be written as,

\[
\dot{u}(t)=ax(t)\times f\left(x(t-\tau_{1}),x(t-\tau_{2})\right),
\]
where,

\begin{equation}
f\left(x(t-\tau_{1}),x(t-\tau_{2})\right)=1-\big(b_{1}x(t-\tau_{1})+b_{2}x(t-\tau_{2})\big).
\end{equation}
Without loss of generality assume that $\tau_{2}\geq\tau_{1}$. (29)
can also be written as,

\begin{equation}
\dot{u}(t)=L_{\mu}u_{t}+F(u_{t},\mu),
\end{equation}
$u(t)=x(t)$, $t>0$, $\mu\in R$, where for $\tau_{2}>0,$

\[
u_{t}(\theta)=u(t+\theta),u:[-\tau_{2},0]\rightarrow R,\theta\in[-\tau_{2},0].
\]
Also, $L_{\mu}:C[-\tau_{2},0]\rightarrow R$ is

\[
L_{\mu}\phi=-(a_{0}+\mu)x^{*}\big(b_{1}\phi(-\tau_{1})+b_{2}\phi(-\tau_{2})\big)
\]
and $F(u_{t},\mu):C[-\tau_{2},0]\rightarrow R$ is 

\[
F(u_{t},\mu)=-(a_{0}+\mu)x^{*}\left(b_{1}u_{t}(0)u_{t}(-\tau_{1})+b_{2}u_{t}(0)u_{t}(-\tau_{2})\right).
\]
Here $a=a_{0}+\mu$ and $a_{0}$ refers to the critical value of the
growth rate at the bifurcation. 

By the Riesz representation theorem, there exists a matrix-valued
function with bounded variation components $\eta(\theta,\mu),\theta\in[-\tau_{2},0]$,
such that

\begin{equation}
L_{\mu}\phi=\int_{-\tau_{2}}^{0}d\eta(\theta,\mu)\phi(\theta),
\end{equation}
where $\phi\in C\left([-\tau_{2},0],\Re\right)$ and,

\[
d\eta(\theta,\mu)=-(a_{0}+\mu)\times x^{*}\big[b_{1}\delta(\theta+\tau_{1})+b_{2}\delta(\theta+\tau_{2})\big].
\]
Here $\delta(\theta)$ is the Dirac-Delta function. 

For $\phi\in C^{1}\left([-\tau_{2},0],\Re\right),$ define

\begin{equation}
A_{\mu}\phi(\theta)=\begin{cases}
\frac{d\phi}{d\theta}, & \theta\in[-\tau_{2},0)\\
\int_{-\tau_{2}}^{0}d\eta(\xi,\mu)\phi(\xi)=L_{\mu}\phi, & \theta=0,
\end{cases}
\end{equation}
and

\begin{equation}
R=\begin{cases}
0 & \theta\in[-\tau_{2},0),\\
F & \theta=0.
\end{cases}
\end{equation}
Now, ($30$) can be written as

\begin{equation}
\dot{u}_{t}(t)=A_{\mu}u_{t}+R_{\mu}u_{t}.
\end{equation}
The bifurcating solutions of $u(t,\mu(\epsilon))$ of ($29$) have
amplitude $O(\epsilon)$, period $P(\epsilon)$ and non-zero Floquet
exponent $\beta(\epsilon)$, where $\mu$, $P$ and $\beta$ have
the following expansions:

\[
\mu=\mu_{2}\epsilon^{2}+\mu_{4}\epsilon^{4}+O(\epsilon^{6}),
\]

\[
P=4\tau_{1}\left(1+T_{2}\epsilon^{2}+T_{4}\epsilon^{4}+O(\epsilon^{6})\right),
\]

\[
\beta=\beta_{2}\epsilon^{2}+\beta_{4}\epsilon^{4}+O(\epsilon^{6}).
\]

If $\mu_{2}>0$, the bifurcation is supercritical and if $\mu_{2}<0,$
it is subcritical. If $\beta_{2}<0$, $u\left(t,\mu(\epsilon)\right)$
shows asymptotic orbital stability and instability if $\beta_{2}>0$.
These coefficients will now be calculated.

Define the adjoint operator $A^{*}$ as,

\begin{equation}
A_{0}^{*}\psi(s)=\begin{cases}
-\frac{d\psi}{ds}, & s\in(0,\tau_{1}]\\
\int_{-\tau_{2}}^{0}d\eta^{T}(t,0)\psi(-t), & s=0.
\end{cases}
\end{equation}
Here, $\psi\in C\left([0,\tau_{2}],\Re\right)$and $\eta^{T}$ denotes
the transpose of $\eta$. 

For $\phi\in C^{1}\left([-\tau_{2},0],\Re\right)$ and $\psi\in C\left([0,\tau_{2}],\Re\right)$
define an inner product

\begin{equation}
\left\langle \psi,\phi\right\rangle =\bar{\psi}(0)\phi(0)-\int_{\theta=-\tau_{2}}^{0}\int_{\xi=0}^{\theta}\bar{\psi}^{T}(\xi-\theta)d\eta(\theta,0)\phi(\xi)d\xi.
\end{equation}
Let $q(\theta)$ be the eigenfunction for $A_{0}$ corresponding to
$\lambda(0)$, namely

\[
A_{0}q(\theta)=j\omega_{0}q(\theta),
\]

\[
q(\theta)=e^{j\omega_{0}\theta}.
\]
Let $q^{*}(s)$ be an eigen vector of $A_{0}^{*}$ such that

\[
q^{*}(s)=De^{j\omega_{0}s},
\]
and 

\[
\left\langle q^{*},q\right\rangle =1,\quad\left\langle q^{*},\bar{q}\right\rangle =0.
\]
From the above equation $D$ can be determined as shown below.

\[
\left\langle q^{*},q\right\rangle =\bar{D}-\bar{D}\int_{\theta=-\tau_{2}}^{0}\int_{\xi=0}^{\theta}e^{-j\omega_{0}(\xi-\theta)}d\eta(\theta)e^{j\omega_{0}\xi}d\xi
\]

\begin{multline*}
\Rightarrow1=\bar{D}-\bar{D}\underset{-\tau_{2}}{\overset{0}{\int}}\theta e^{j\omega_{0}\theta}d\eta(\theta)\\
\Rightarrow1=\bar{D}-\bar{Dx^{*}}\left[\tau_{1}e^{-j\omega_{0}\tau_{1}}(a_{0}+\mu)b_{1}+\tau_{2}e^{-j\omega_{0}\tau_{2}}(a_{0}+\mu)b_{2}\right]\\
\Rightarrow D=\frac{1}{1-(a_{0}+\mu)x^{*}\left[b_{1}\tau_{1}e^{j\omega_{0}\tau_{1}}+b_{2}\tau_{2}e^{j\omega_{0}\tau_{2}}\right]}.
\end{multline*}
It can be easily verified that $\left\langle q^{*},\bar{q}\right\rangle =0$.

Define,

\[
z(t)=\left\langle q^{*},u_{t}\right\rangle ,
\]
\[
w(t,\theta)=u_{t}(\theta)-2\Re\left\{ z(t)q(\theta)\right\} .
\]
Then on the centre manifold $C_{0}$, 

\[
w(t,\theta)=w\left(z(t),\bar{z}(t),\theta\right),
\]
where

\[
w(z,\bar{z},\theta)=w_{20}(\theta)\frac{z^{2}}{2}+w_{11}(\theta)z\bar{z}+w_{02}(\theta)\frac{\bar{z}^{2}}{2}+...
\]
and $z$ and $\bar{z}$ are local coordinates for centre manifold
$C_{0}$ in the direction of $q^{*}$ and $\bar{q}^{*}$. Note that
$w$ is also real if $u_{t}$ is real, we consider only real solutions.
For solutions $u_{t}\in C_{0}$ of ($11$) at $\mu=0$,

\begin{flalign}
\dot{z}(t) & =\left\langle q^{*},Au_{t}+Ru_{t}\right\rangle \nonumber \\
 & =j\omega_{0}z(t)+\bar{q}^{*}(0)F_{0}(z,\bar{z})\nonumber \\
 & =j\omega_{0}z(t)+g(z,\bar{z)},
\end{flalign}
here,
\[
g(z,\bar{z})=\bar{q}^{*}(0)F_{0}(z,\bar{z})=g_{20}\frac{z^{2}}{2}+g_{11}z\bar{z}+g_{02}\frac{\bar{z}^{2}}{2}+g_{21}\frac{z^{2}\bar{z}}{2}+...
\]
Now using ($34$) and ($35$) we get,

\begin{equation}
\dot{w}=\dot{u_{t}}-\dot{z}q-\dot{\bar{z}}\bar{q}
\end{equation}
or,

\[
\dot{w}=\begin{cases}
Aw-2\Re\left\{ \bar{q}^{*}(0)F_{0}q(\theta)\right\} , & \theta\in[-\tau_{1},0)\\
Aw-2\Re\left\{ \bar{q}^{*}(0)F_{0}q(\theta)\right\} +F_{0}, & \theta=0,
\end{cases}
\]
which can be written as

\begin{equation}
\dot{w}=Aw+H(z,\bar{z},\theta),
\end{equation}
where

\begin{equation}
H(z,\bar{z},\theta)=H_{20}(\theta)\frac{z^{2}}{2}+H_{11}(\theta)z\bar{z}+H_{02}(\theta)\frac{\bar{z}^{2}}{2}+...
\end{equation}
Expanding the above series and comparing the co-efficients we get,

\begin{equation}
(2j\omega_{0}-A)w_{20}(\theta)=H_{20}(\theta),
\end{equation}

\begin{equation}
-Aw_{11}=H_{11}(\theta),
\end{equation}

\begin{equation}
-(2j\omega_{0}+A)w_{02}(\theta)=H_{02}(\theta).
\end{equation}
From ($38$), we have

\begin{multline*}
u_{t}(\theta)=w(z,\bar{z},\theta)+zq(\theta)+\bar{z}\bar{q}(\theta)\\
=w_{20}(\theta)\frac{z^{2}}{2}+w_{11}(\theta)z\bar{z}+w_{02}(\theta)\frac{\bar{z}^{2}}{2}+ze^{j\omega_{0}\theta}+\bar{z}e^{-j\omega_{0}\theta}+...
\end{multline*}
from which $u_{t}(0)$ and $u_{t}(-\tau_{1})$ can be obtained. As
we only need the coefficients of $z^{2},$ $z\bar{z}$, $z^{2}$ and
$z^{2}\bar{z}$, we get

\[
u_{t}(0)=w(z,\bar{z},0)+z+\bar{z},
\]

\[
u_{t}(-\tau_{1})=w(z,\bar{z},-\tau_{1})+ze^{-j\omega_{0}\tau_{1}}+\bar{z}e^{j\omega_{0}\tau_{1}},
\]

\begin{eqnarray*}
u_{t}(0)u_{t}(-\tau_{1}) & = & w(0)w(-\tau_{1})+w(-\tau_{1})(z+\bar{z})\\
 &  & +w(0)\left(ze^{-j\omega_{0}\tau_{1}}+\bar{z}e^{j\omega_{0}\tau_{1}}\right)\\
 &  & +z^{2}e^{-j\omega_{0}\tau_{1}}+z\bar{z}\left(e^{j\omega_{0}\tau_{1}}+e^{-j\omega_{0}\tau_{1}}\right)+\bar{z}^{2}e^{j\omega_{0}\tau_{1}},
\end{eqnarray*}

\begin{eqnarray*}
u_{t}(0)u_{t}(-\tau) & = & z^{2}e^{-j\omega_{0}\tau_{1}}+\bar{z}^{2}e^{j\omega_{0}\tau_{1}}+z\bar{z}(e^{j\omega_{0}\tau_{1}}+e^{-j\omega_{0}\tau_{1}})\\
 &  & +z^{2}\bar{z}(2w_{11}(0)e^{-j\omega_{0}\tau_{1}}+\frac{w_{20}(0)}{2}e^{-j\omega_{0}\tau_{1}}\\
 &  & +w_{11}(-\tau)+\frac{w_{20}(-\tau)}{2})+...
\end{eqnarray*}
Now we can write,

\begin{eqnarray*}
g(z,\bar{z}) & = & \bar{q}^{*}(0)F_{0}(z,\bar{z})\\
 & = & g_{20}\frac{z^{2}}{2}+g_{11}z\bar{z}+g_{02}\frac{\bar{z}^{2}}{2}+g_{21}\frac{z^{2}\bar{z}}{2}+...
\end{eqnarray*}

\begin{eqnarray}
g_{20} & = & \bar{-q^{*}}(0)2ax^{*}\left[b_{1}e^{j\omega_{0}\tau_{1}}+b_{2}e^{j\omega_{0}\tau_{2}}\right]\nonumber \\
 & = & -\bar{D}2ax^{*}\left[b_{1}e^{-j\omega_{0}\tau_{1}}+b_{2}e^{-j\omega_{0}\tau_{2}}\right]
\end{eqnarray}

\begin{equation}
g_{11}=-\bar{D}ax^{*}\left[b_{1}(e^{j\omega_{0}\tau_{1}}+e^{-j\omega_{0}\tau_{1}})+b_{2}(e^{j\omega_{0}\tau_{2}}+e^{-j\omega_{0}\tau_{2}})\right],
\end{equation}

\begin{equation}
g_{02}=-\bar{D}2ax^{*}[b_{1}e^{j\omega_{0}\tau_{1}}+b_{2}e^{j\omega_{0}\tau_{2}}],
\end{equation}

\begin{eqnarray*}
g_{21} & = & -\bar{D}2ax^{*}\bigg\{ b\bigg(w_{11}(0)e^{-j\omega_{0}\tau_{1}}+\frac{w_{20}(0)}{2}e^{j\omega_{0}\tau_{1}}\\
 &  & +w_{11}(-\tau_{1})+\frac{w_{20}(-\tau_{1})}{2}\bigg)+b_{2}\bigg(w_{11}(0)e^{-j\omega_{0}\tau_{2}}\\
 &  & +\frac{w_{20}(0)}{2}e^{j\omega_{0}\tau_{2}}+w_{11}(-\tau_{2})+\frac{w_{20}(-\tau_{2})}{2}\bigg)\bigg\}
\end{eqnarray*}
We have, for $\theta\in[-\tau_{1},0)$, 

\begin{eqnarray*}
H(z,\bar{z},\theta) & = & 2\Re{\bar{z}^{*}(0)F_{0}q(\theta)}\\
 & = & -gq(\theta)-\bar{g}\bar{q}(\theta)\\
 & = & -\left(g_{20}\frac{z^{2}}{2}+g_{11}z\bar{z}+g_{02}\frac{\bar{z}^{2}}{2}+...\right)q(\theta)\\
 & = & -\left(g_{20}\frac{z^{2}}{2}+g_{11}z\bar{z}+g_{02}\frac{\bar{z}^{2}}{2}+...\right)\bar{q}(\theta)
\end{eqnarray*}
which yields

\[
H_{20}=-g_{20}q(\theta)-\bar{g}_{02}q(\theta),
\]

\[
H_{11}=-g_{11}q(\theta)-\bar{g}_{11}q(\theta).
\]
Using the above in ($41$) to ($43$), we get

\begin{equation}
\dot{w}_{20}(\theta)=2j\omega_{0}w_{20}(\theta)+g_{20}q(\theta)+\bar{g}_{02}\bar{q}(\theta)
\end{equation}

\begin{equation}
\dot{w}_{11}(\theta)=g_{11}q(\theta)+\bar{g}_{11}\bar{q}(\theta).
\end{equation}
On solving the above differential equations,

\begin{equation}
w_{20}(\theta)=-\frac{g_{20}}{j\omega_{0}}q(0)e^{j\omega_{0}\theta}-\frac{\bar{g}_{02}}{3j\omega_{0}}\bar{q}(0)e^{-j\omega_{0}\theta}+E_{1}e^{2j\omega_{0}\theta}
\end{equation}

\begin{equation}
w_{11}(\theta)=\frac{g_{11}}{j\omega_{0}}q(0)e^{j\omega_{0}\theta}-\frac{\bar{g}_{11}}{j\omega_{0}}\bar{q}(0)e^{-j\omega_{0}\theta}+E_{2}.
\end{equation}
for some $E_{1},E_{2}$ which will soon be determined.

Similarly, for $\theta=0,$

\[
H(z,\bar{z},0)=-\Re{\bar{q}^{*}F_{0}q(0)}+F_{0}
\]

\begin{eqnarray}
H_{20}(0) & = & -g_{20}q(0)-\bar{g}_{02}\bar{q}(0)-2a\big(b_{1}e^{-j\omega_{0}\tau_{1}}\nonumber \\
 &  & +b_{2}e^{-j\omega_{0}\tau_{2}}\big)\\
 & = & ax^{*}\left(b_{1}w_{20}(-\tau_{1})+b_{2}w_{20}(-\tau_{2})\right)\\
 &  & +2j\omega_{0}w_{20}(0)
\end{eqnarray}

\begin{eqnarray}
H_{11}(0) & = & -g_{11}q(0)-\bar{g}_{11}\bar{q}(0)-2a\bigg(b_{1}(e^{j\omega_{0}\tau_{1}}+e^{-j\omega_{0}\tau_{1}})\nonumber \\
 &  & +b_{2}(e^{j\omega_{0}\tau_{2}}+e^{-j\omega_{0}\tau_{2}})\bigg).\\
 & = & ax^{*}\left(b_{1}w_{11}(-\tau_{1})+b_{2}w_{11}(-\tau_{2})\right)
\end{eqnarray}
The expression for $w_{20}(-\tau_{1})$, $w_{20}(-\tau_{2})$, $w_{20}(0)$,
$w_{11}(-\tau_{1})$, $w_{11}(-\tau_{2})$, and $w_{11}(0)$ can be
found from ($49$) and ($50)$. On substituting them in ($51$) and
($52$), $E_{1},E_{2}$ can be found. From the above analysis the
following can be calculated:

\begin{equation}
c_{1}(0)=\frac{j}{2\omega_{0}}\left(g_{20}g_{11}-2\left|g_{11}\right|^{2}-\frac{1}{3}\left|g_{02}\right|^{2}\right)+\frac{g_{21}}{2},
\end{equation}

\begin{equation}
\mu_{2}=-\frac{\Re(c_{1}(0))}{\Re(\lambda'(a_{0}))},
\end{equation}

\begin{equation}
P=4\tau(1+\epsilon^{2}T_{2}+O(\epsilon^{4})),
\end{equation}

\begin{equation}
T_{2}=-\left(\frac{\Im\big(c_{1}(0)+\mu_{2}\lambda'(a_{0})\big)}{\omega_{0}}\right),
\end{equation}

\begin{equation}
\beta=\epsilon^{2}\beta_{2}+O(\epsilon^{4}),\;\beta_{2}=2\Re(c_{1}(0)),\;\epsilon=\sqrt{\frac{\mu}{\mu_{2}}},
\end{equation}
where $c_{1}(0)$ is the lyapunov coefficient. The asymptotic form
of the bifurcating periodic solutions is

\begin{equation}
u(t,\mu(\epsilon))=2\epsilon\Re\left(q(0)e^{j\omega_{0}t}\right)+\epsilon^{2}\Re\left(E_{1}e^{2j\omega_{0}t}+E_{2}\right)+O(\epsilon^{3})
\end{equation}
for $0\leq t\leq P(\epsilon)$.

On substituting $\tau_{1}=\tau_{2}$ in ($29$) the analysis reduces
to that of the single delay logistic equation.

\subsection{Direction and stability of the Hopf bifurcation in the logistic equation
with delay as the parameter}

In the previous sub-section the direction and stability of the bifurcation
was considered with the growth rate as the bifurcation parameter.
It has been shown in Section 5.1 that the single delay equation undergoes
Hopf bifurcation with an increase in the delay also. A similar analysis
for this case is carried out here. 

Let $\tau_{0}$ be the critical value of the delay at bifurcation.
The logistic equation can also be written as,

\begin{equation}
\dot{u}(t)=L_{\mu}u_{t}+F(u_{t},\mu),
\end{equation}
where, $u(t)=x(\tau_{1}t)$, $t>0$, $\mu\in R$, $\tau_{1}=\tau_{0}+\mu$,
and

\[
u_{t}(\theta)=u(t+\theta)u:[-1,0]\rightarrow R,\theta\in[-1,0].
\]
Also, $L_{\mu}:C[-1,0]\rightarrow R$ is

\begin{equation}
L_{\mu}\phi=-ab\phi(-1).
\end{equation}
$F(u_{t},\mu):C[-1,0]\rightarrow R$ is 

\begin{equation}
F(\phi,\mu)=-ab\phi(0)\phi(-1).
\end{equation}

By the Riesz representation theorem, there exists a matrix function
with bounded variation components $\eta(\theta,\mu),\theta\in[-1,0]$,
such that

\begin{equation}
L_{\mu}\phi=\int_{-1}^{0}d\eta(\theta,\mu)\phi(\theta).
\end{equation}
where,

\[
d\eta(\theta,\mu)=-ab\delta(\theta+1)d\theta,
\]
where $\delta(\theta)$is the Dirac delta function.

Proceeding further as shown in the previous sub-section we get,

\begin{equation}
\mu_{2}=\frac{(3\pi-2)}{10}>0.
\end{equation}
Hence the Hopf bifurcation is supercritical.

\subsubsection*{Numerical Example}

If we take the case of a system with a two delays with $b_{1}=b_{2}=0.5$,
$\tau_{1}=1$ and $\tau_{2}=2$, the system undergoes a Hopf bifurcation
at growth parameter $a_{0}=1.2092$. The value of the lyapunov coefficient
$c_{1}(0)=-0.1691-0.2290i,$ the real part of which is less than zero
which makes the resulting periodic solutions asymptotically orbitally
stable. Parameter $\mu_{2}=0.5175>0,$ implying that the Hopf bifurcation
is supercritical. The period of these oscillations is $P=6.$ Fig.
$10$ has the resulting state space diagram where the limit cycles
can be seen and Fig. $11$ compares the bifurcation diagram obtained
through simulation to the analytical solution. 

\begin{figure}
\includegraphics[width=8cm]{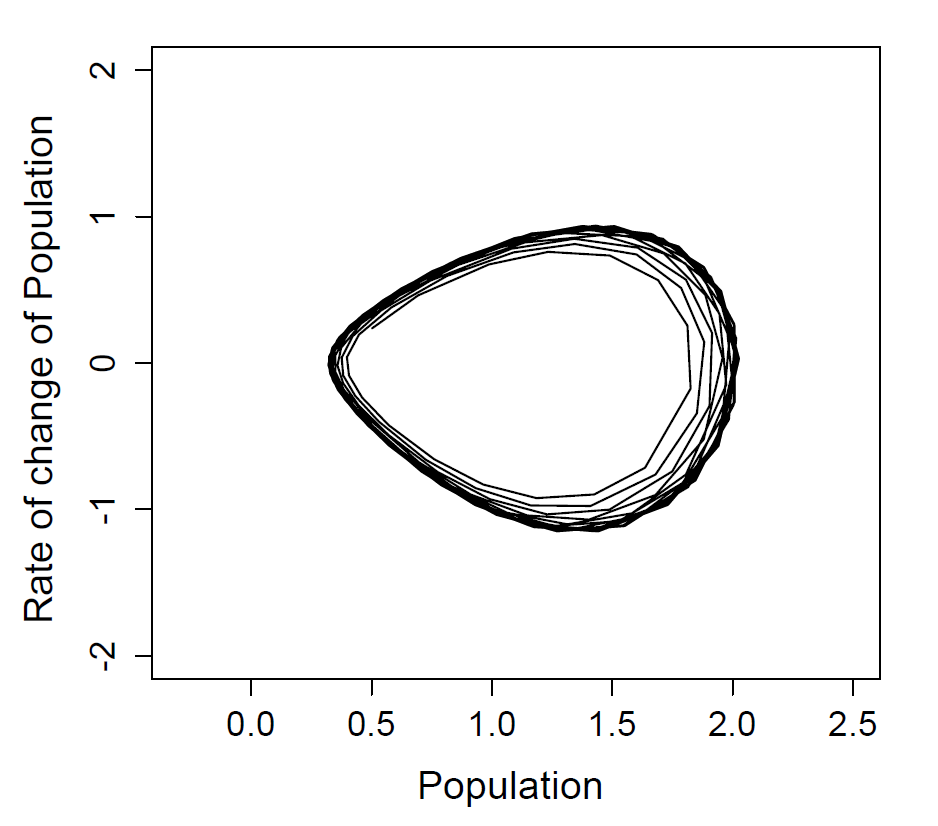}

\caption{The state space diagram for the two delay system as shown in the numerical
example with the value of $a=a_{0}+0.05.$}
\end{figure}

\begin{figure}
\includegraphics[width=8cm]{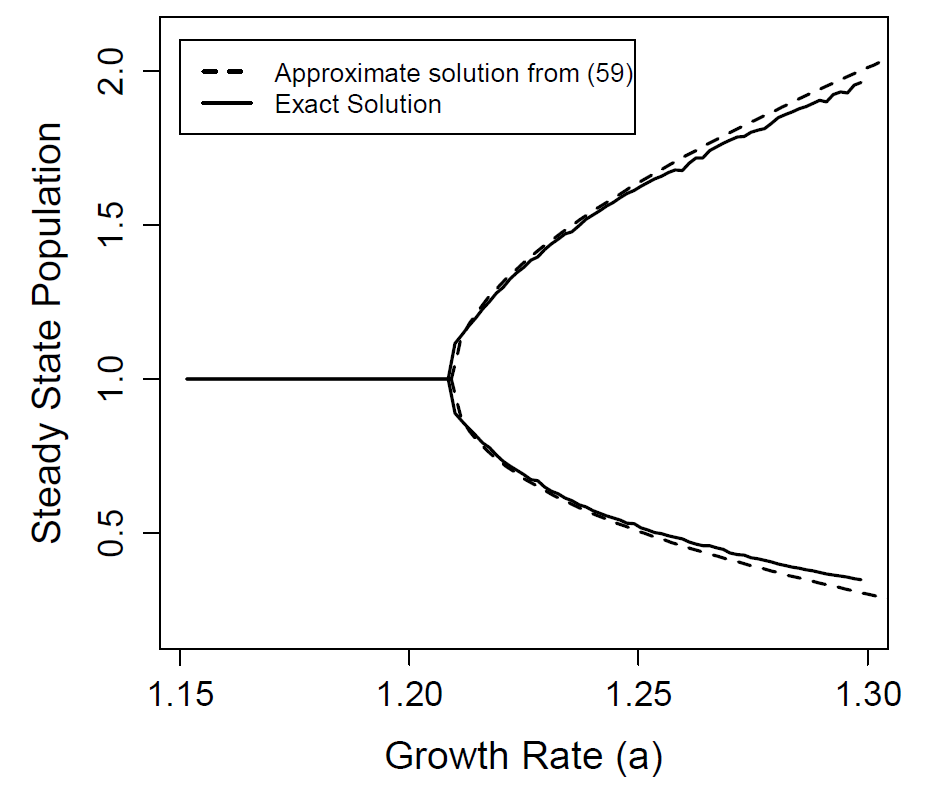}

\caption{The bifurcation diagram for the system with two delays as described
in the numerical example. As can be seen, the results obtained through
simulation match the analytical predictions for values of the bifurcation
parameter $\mu=a-a_{0}<0.1.$}
\end{figure}

\subsection{Bifurcation diagrams}

In Fig. $12$ and $13$, the bifurcation diagrams of the single delay
system are drawn with respect to both parameters namely, the growth
rate $a$ and delay $\tau_{1}.$ Both curves look similar. In Fig.
$14$ and $15$, the bifurcation diagrams of the two delay system
are shown with respect to various parameters. The only parameter that
affects the equilibrium value is the scaling factor of the delay terms,
$b_{1}$ and $b_{2}.$ The amplitude of the periodic oscillations
beyond the critical value of the parameter varies more with changes
in the growth rate $a$ than it does with delays $\tau_{1}$ and $\tau_{2}.$ 

\begin{figure}
\includegraphics[width=8cm]{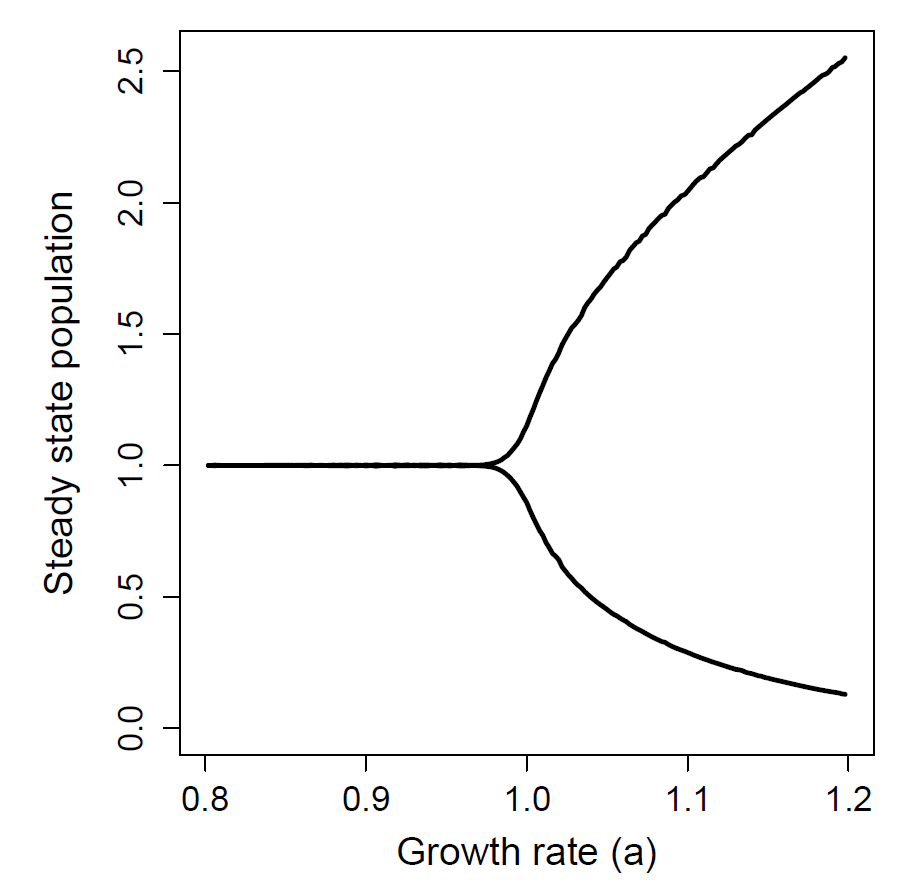}

\caption{The bifurcation diagram of a single delay system with respect to the
growth rate $a$. Here the value of the delay $\tau_{1}=\frac{\pi}{2}.$
As can be evidenced from the graph, the equilibrium value is constant
with respect to $a$ before the critical value $a=1$. The amplitude
increases sharply with changes in $a.$}
\end{figure}

\begin{figure}
\includegraphics[width=8cm]{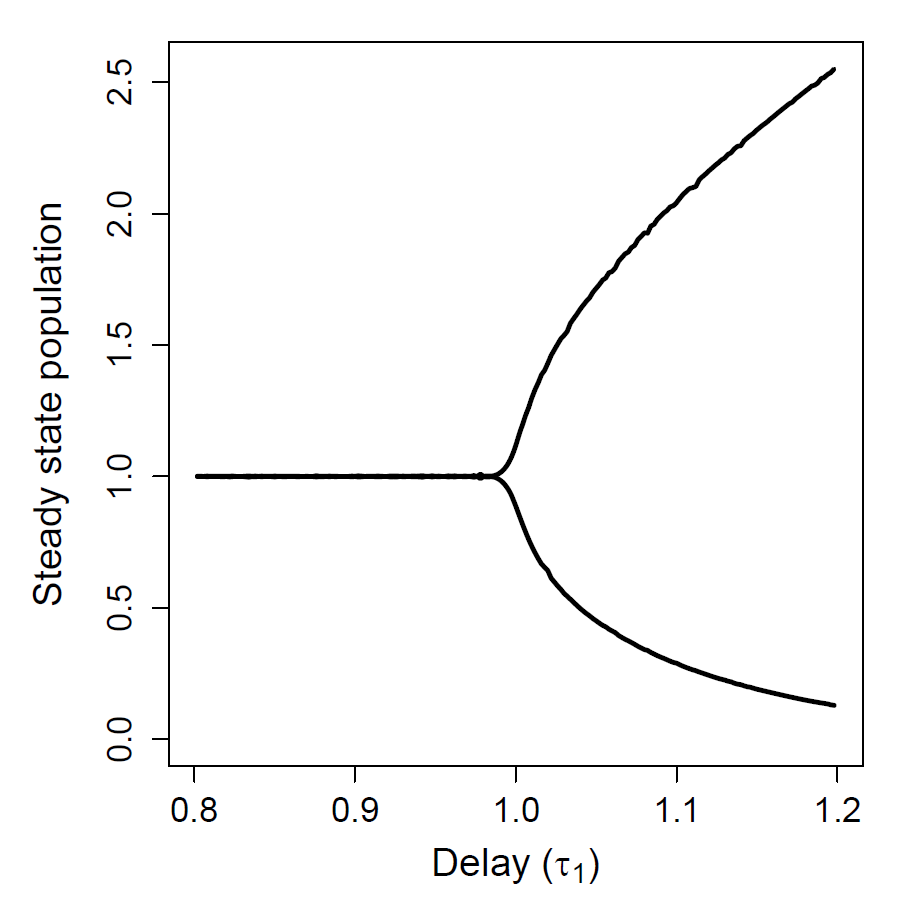}

\caption{The bifurcation diagram of a single delay system with respect to the
time delay $\tau_{1}$ while the growth rate $a=\frac{\pi}{2}$. The
equilibrium point is not dependent on $\tau_{1}$ when the system
is stable and the amplitude of the periodic oscillations vary as much
with variations in $\tau_{1}$ beyond the critical value $\tau_{1}=1$
as it did with variations in $a$. }
\end{figure}

\begin{figure}
\includegraphics[width=8cm]{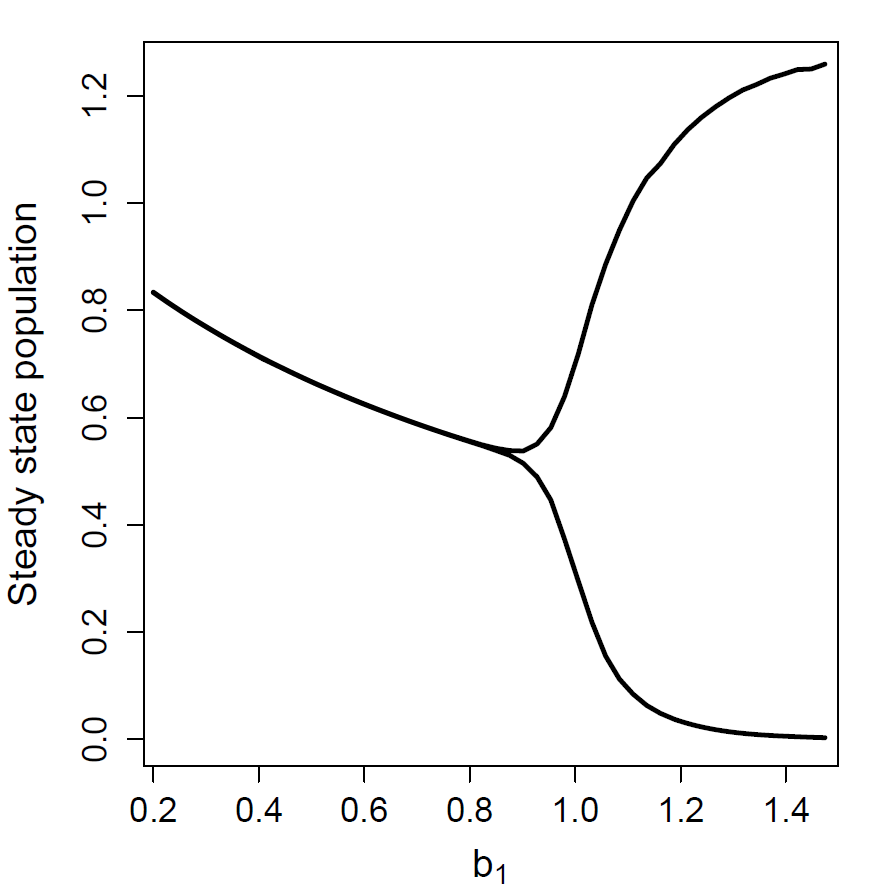}

\caption{The bifurcation diagram of a two delay system with variations in $b_{2}$
when $a=b_{1}=\tau_{1}=1$ and $\tau_{2}=10$. The equilibrium value
depends on this parameter. The amplitude of periodic oscillations
does not vary as sharply with variations in $b_{2}$ beyond the critical
value. }
\end{figure}

\begin{figure}
\includegraphics[width=8cm]{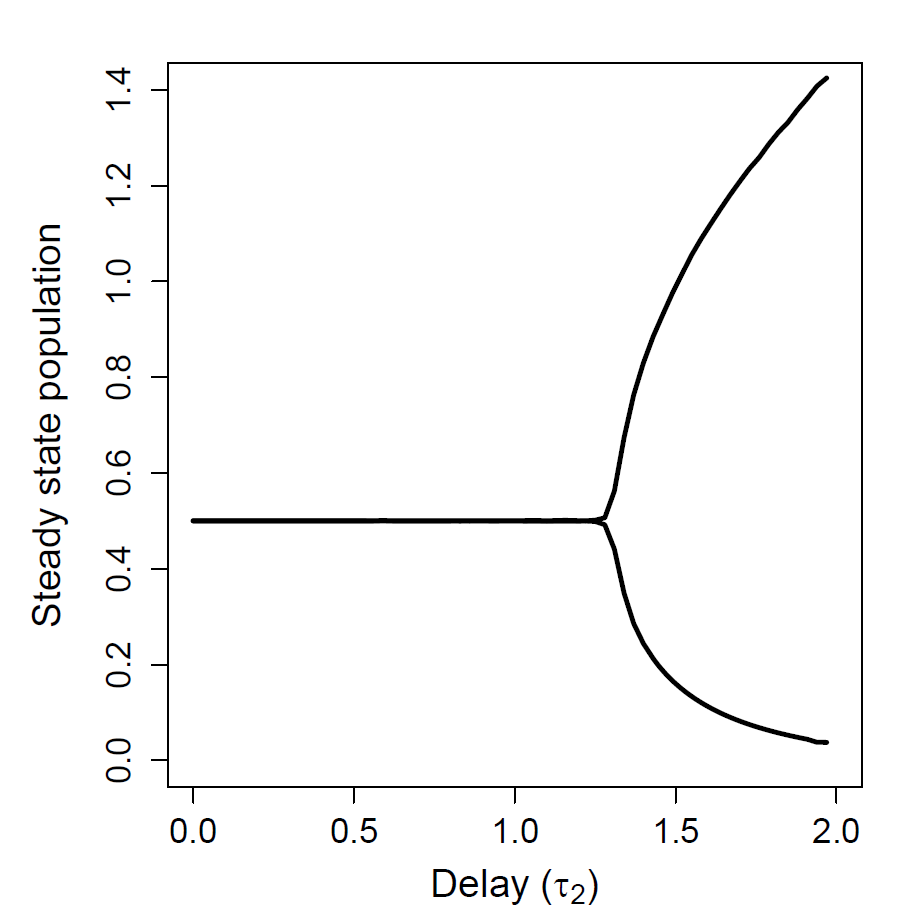}

\caption{The bifurcation diagram of the two delay system with parameter $\tau_{2}$
when $a=b_{1}=b_{2}=1$ and $\tau_{1}=2.$ The variation of amplitude
of periodic solution is gradual with variations of $\tau_{2}$ beyond
the critical value.}
\end{figure}

\section{Results}

In this section, we show the behaviour of the delay logistic system
with time under various conditions. In Fig. $16$, we see the sigmoid
curve of the logistic function. As growth rate is increased, the function
reaches saturation value faster. In Fig. $17$, the time domain response
of a converging system is observed. In Fig. $18$, the logistic equation
with two delays is observed in the time domain. Varying the parameters
changes the behaviour from asymptotically stable to forming limit
cycles. In the latter case, the sufficient and necessary conditions
for stability are not met. 

\begin{figure}
\includegraphics[width=8cm]{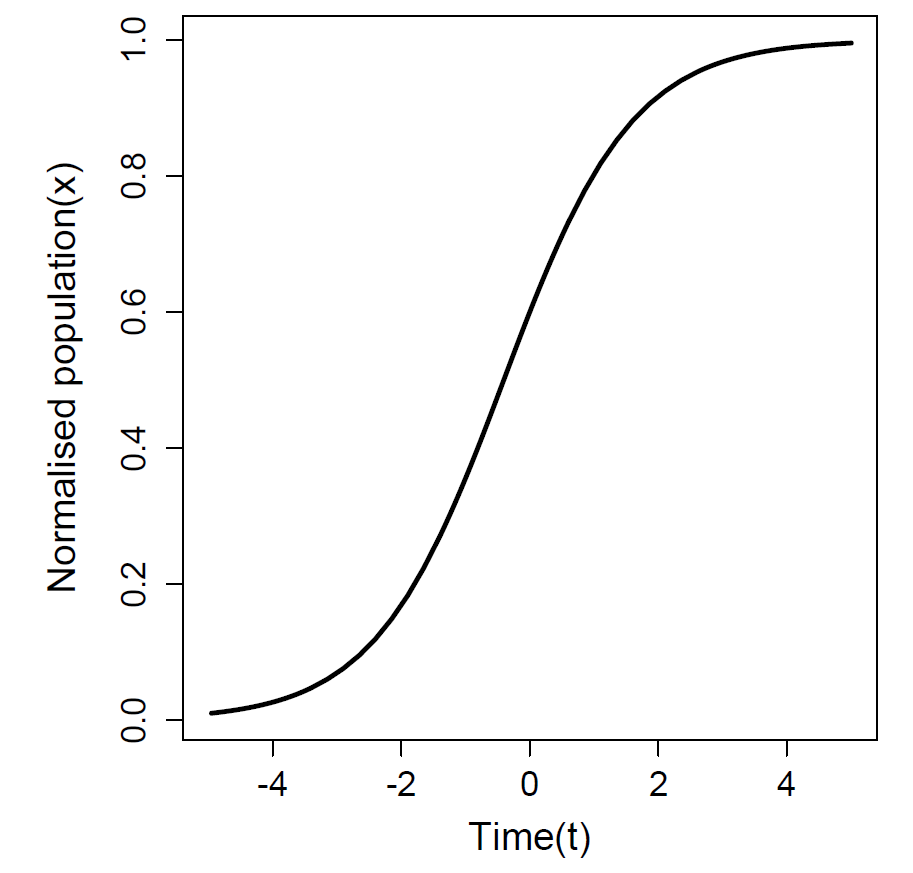}

\caption{Time domain response of logistic equation with no delay ($a=1$).
Note the sigmoid shape of the curve. The initial growth rate is exponential
but as the population approaches the maximum carrying capacity, it
saturates due to diminished resources. The logistic equation was a
correction to Robert Malthus' proposition that population growth is
purely exponential. }
\end{figure}

\begin{figure}
\includegraphics[width=8cm]{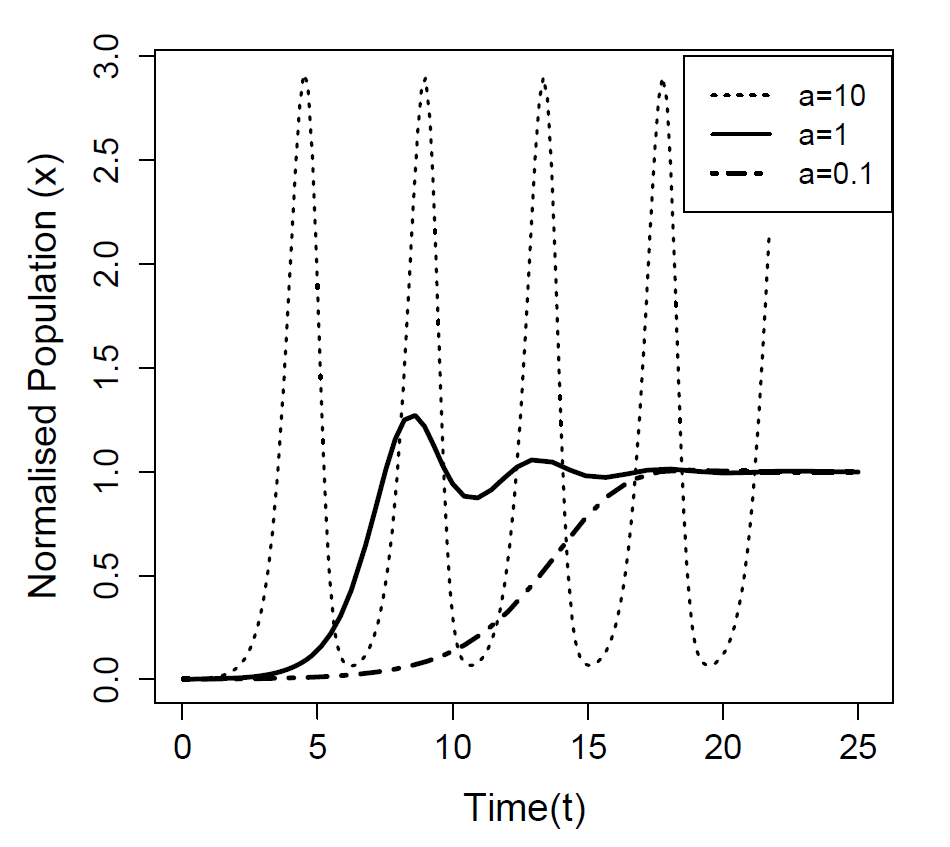}

\caption{Varying coefficient of the growth rate term $a$ at constant value
of delay $\tau_{1}=1$. This may be viewed as the population charts
of different species, each with a specific resource consumption rate
per capita. The species that consume more are the ones that saturate
at a lower population. Also, the behaviour of a stable and converging
system can be seen.}
\end{figure}

\begin{figure}
\includegraphics[width=8cm]{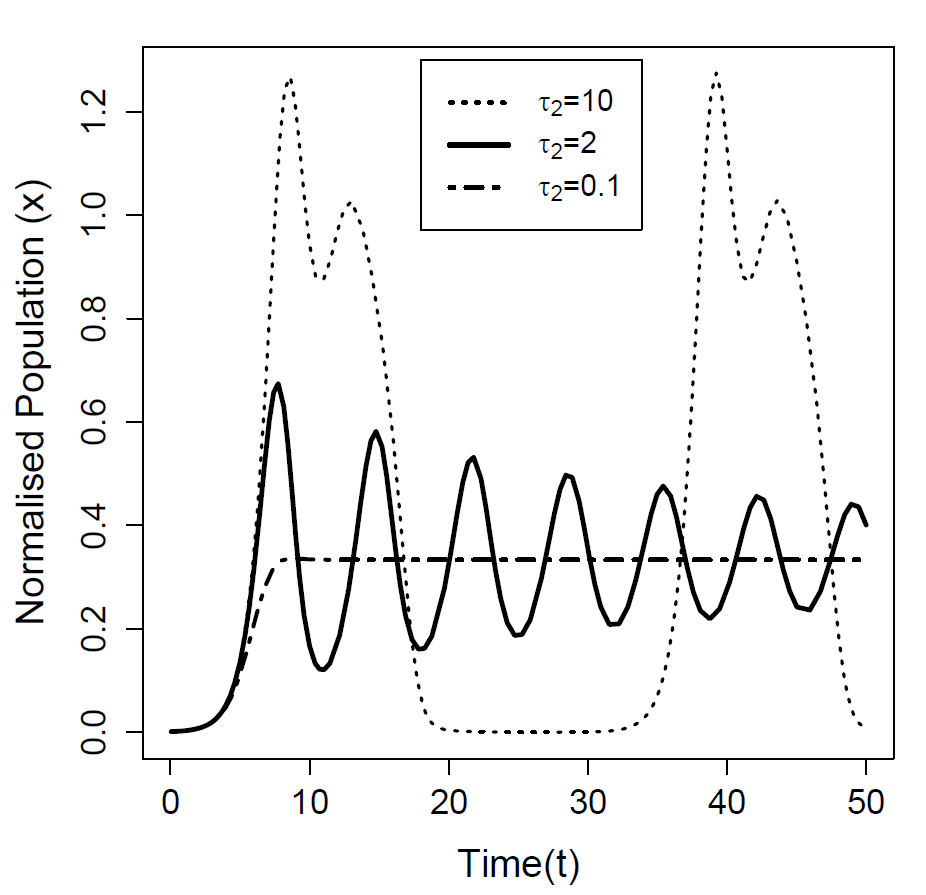}

\caption{Varying $\tau_{2}$ at $\tau_{1}=1$, second delay coefficient being
twice the first. i.e. $b_{2}=2b_{1}$. When $\tau_{2}=2$ or lower,
the system is converging to an equilibrium point. When $\tau_{2}=10$,
limit cycles are seen. }
\end{figure}

\section{Conclusions}

We have performed local stability analysis of the logistic equation
with and without multiple time delays. Sufficient conditions to aid
design were also extracted. The rate of convergence for the single
delay logistic equation was analysed. While the ideal system (without
delay) is always perfectly stable, the actual system that we have
considered using delays undergoes a Hopf bifurcation which is supercritical.
The nature of the resulting periodic oscillations was analytically
characterized for the single delay system and a methodology to ascertain
the same in the two delay case was presented.

\section*{Acknowledgements}

We extend our gratitude to Prof. Gaurav Raina for his guidance, insight
and invaluable suggestions.

\end{document}